\documentclass[floatfix,twocolumn,showpacs,aps,prb,amsmath,superscriptaddress]{revtex4}
\usepackage{graphicx}
\usepackage{bm}
\usepackage{amssymb}
\usepackage{dcolumn}
\usepackage{hyperref}
\hypersetup{
		colorlinks=true,
		linkcolor=magenta,
		citecolor=blue,
		urlcolor=blue,
		pdfstartview={FitH},
		}

\newcommand{\q}[1]{\mathrm{#1}}
\newcommand{\mf}[1]{\ensuremath{\boldsymbol{\q{#1}}}}

\newcommand{\kf}{{\mf k}}

\newcommand{\jf}{\mf j}

\newcommand{\bracks}[1]{\ensuremath{\left[ #1 \right]}}

\newcommand{\vareps}{\varepsilon}

\newcommand{\bege}{\begin{equation}\begin{aligned}}
\newcommand{\ee}{\end{aligned}\end{equation}}

\newcommand{\sigmaic}{\ensuremath{\sigma^{\q{ic}}}}
\newcommand{\sigmasj}{\ensuremath{\sigma^{\q{sj}}}}
\newcommand{\alphaic}{\ensuremath{\alpha^{\q{ic}}}}
\newcommand{\alphasj}{\ensuremath{\alpha^{\q{sj}}}}

\newcommand{\fig}[1]{Fig.~(\ref{#1})}
\newcommand{\eq}[1]{Eq.~(\ref{#1})}
\newcommand{\tab}[1]{Tab.~(\ref{#1})}

\begin{document}

\title{Scattering-Independent Anomalous Nernst Effect in Ferromagnets}

\author{J\"urgen Weischenberg}
\email[Corresp.~author:~]{j.weischenberg@fz-juelich.de}

\author{Frank Freimuth}

\author{Stefan Bl\"ugel}

\author{Yuriy Mokrousov}
\affiliation{Peter Gr\"unberg Institut and Institute for Advanced Simulation,
Forschungszentrum J\"ulich and JARA, 52425 J\"ulich, Germany}

\date{\today}

\begin{abstract}

Using the full-potential linearized augmented plane-wave method within the
density functional theory, we compute all contributions to the scattering
independent part of the thermoelectric conductivity tensor $\hat\alpha$,
namely the intrinsic contribution $\hat\alpha^{\q{ic}}$ and the side-jump
contribution $\hat\alpha^{\q{sj}}$.  For the ferromagnetic materials bcc Fe, hcp Co,
fcc Ni and $L1_{0}$ ordered alloys FePd and FePt, our investigations of the
energy and temperature dependence of the intrinsic and side-jump
contributions show that they are both of equal importance. Overall, our
calculations are able to correctly reproduce the order of magnitude and sign of
the experimentally measured signal, suggesting that the scattering independent
part plays an important role in the anomalous Nernst effect of ferromagnets.

\end{abstract}


\maketitle

When a thermal gradient is applied to a 
single-domain ferromagnetic material,
the anomalous Nernst effect (ANE) shows itself as an electric
field $\mf E$ that emerges in a direction perpendicular to the applied thermal
gradient $-\boldsymbol\nabla T$ and perpendicular to the sample's magnetization
$\mf M$.\cite{Nernst:1887}  Prominently, the Nernst signal can be used as a
probing tool for the vortex phase in type II
superconductors\cite{Xu:2000,Wang:2001} and it has been also discussed in
spinel ferromagnets\cite{Lee:2004} and on a surface of a topological
insulator.\cite{Yokoyama:2011} However, despite increasing interest in this
phenomenon in the past years,\cite{Bergman:2010} no attempt to predict from
first principles the values of the ANE in metallic ferromagnets, such as bcc
Fe,  has been made so far.

It is convenient to quantify the ANE in terms of the thermoelectric
conductivity tensor $\hat\alpha$, also called the Peltier or Nernst conductivity
tensor.  In linear response theory, the expression for the charge current $\jf$
in the presence of a weak electric field and a small thermal gradient reads
\cite{Luttinger:1964}
\bege
\jf=\hat\sigma\cdot\mf E-\hat\alpha\cdot\boldsymbol\nabla T,
\label{eq:j}
\ee
where the electric conductivity tensor is denoted by $\hat\sigma$. 
The matrix elements of $\hat\alpha$ and $\hat\sigma$ are related via the 
generalized Mott formula\cite{Smrcka:1977,Jonson:1980,Kearney:1988}
\bege
\hat\alpha=-\frac{1}{e}\int d\vareps\,\frac{\partial f }{\partial\mu}
\,\hat\sigma\,\frac{\vareps-\mu}{T},
\label{eq:Mott}
\ee
where $e=-|e|$ is the electronic charge, $\vareps$ the energy  and $\mu$ the
chemical potential of the electrons which appears in the Fermi distribution
function $f(\vareps,\mu,T)$.  In ferromagnetic materials, we can replace the
off-diagonal matrix elements of the conductivity tensor
with the transverse anomalous Hall
conductivity $\sigma^{\q{AHE}}$,\cite{Pugh:1930} which is usually
decomposed into an intrinsic Berry curvature driven contribution
$\sigma^{\q{ic}}$,\cite{Luttinger:1954} and two extrinsic, stemming from
disorder, contributions. Of the latter two, the first is the so-called
side-jump contribution $\sigma^{\q{sj}}$, which is caused by the
scattering of electrons off impurities but which paradoxically does not depend
on their concentration  $n_{i}$.\cite{Berger:1970} The second disorder-driven
contribution is the skew-scattering $\sigma^{\q{sk}}$, which is inversely
proportional to the impurity concentration, i.e., $\sigma^{\q{sk}}\propto
1/n_{i}$.\cite{Smit:1955,Smit:1958} The Mott relation,~\eq{eq:Mott}, implies
that the thermoelectric conductivity tensor can be decomposed in an analogous
way:
\bege
\hat\alpha=\hat\alpha^{\q{ic}}+\hat\alpha^{\q{sj}}+\hat\alpha^{\q{sk}}.
\ee 

The difficulties in understanding the origins of the a\-nom\-a\-lous Hall
effect (AHE) have greatly impeded the progress in the field of the ANE as well.
For example, on the side of qualitative theory of solids, we are aware of only
a single estimate for the value of $\hat\alpha^{\q{ic}}$ in the cuprate
CuCr$_{2}$Se$_{4-x}$Br$_{x}$.\cite{Xiao:2006} However, as we have recently shown,
all scattering in\-de\-pend\-ent contributions to the AHE, that is,
$\hat\sigma^{\q{ic}}$ and $\hat\sigma^{\q{sj}}$, can be calculated from first
principles on an equal footing from the knowledge of the electronic structure
of the pristine crystal alone.\cite{Weischenberg:2011}
Since  skew scattering is suppressed for metals outside the extremely pure regime,
\cite{Miyasato:2007,Nagaosa:2010}
the calculated values for 
$\hat\sigma^{\q{ic}}$ and $\hat\sigma^{\q{sj}}$
allow for a quantitative comparison between theory and experiment.
In the present work, we extend this methodology to the ANE.
We calculate all the
scat\-ter\-ing-in\-de\-pen\-dent contributions to the thermoelectric
conductivity tensor, $\hat{\alpha}^{\q{ic}}$ and $\hat{\alpha}^{\q{sj}}$, in bcc Fe, hcp Co,
fcc Ni and $L1_{0}$ ordered alloys FePd and FePt.  By comparison to
experimental data, we show that $\hat{\alpha}^{\q{ic}}$ and $\hat{\alpha}^{\q{sj}}$ provide the
correct order of magnitude and sign of the anomalous Nernst signal in
transition-metal ferromagnets. We also make predictions concerning the
temperature dependence of the scat\-ter\-ing-in\-de\-pend\-ent ANE.

\begin{figure*}[ht]
\includegraphics[width=\textwidth]{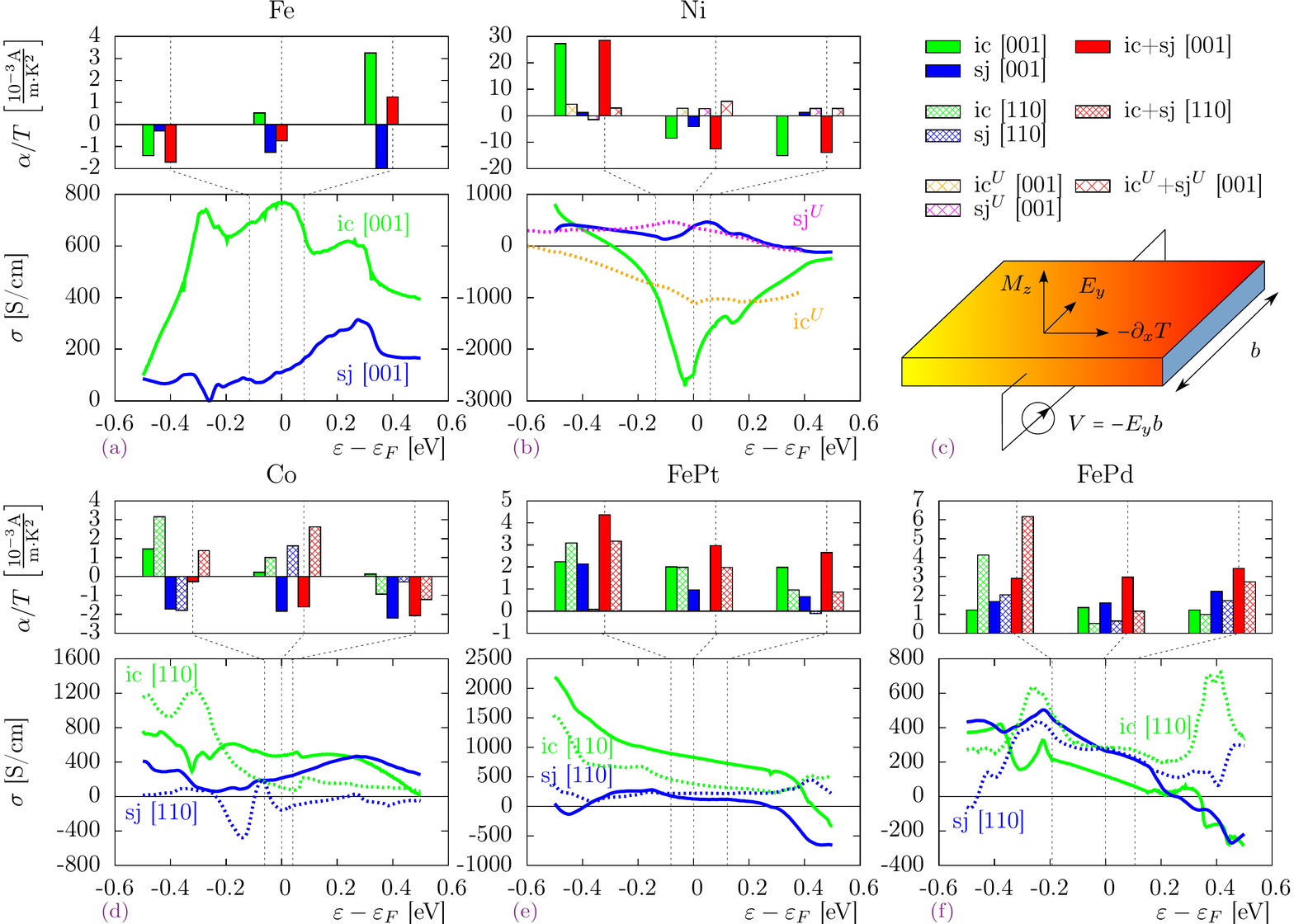}
\caption{(Color online) ANE at  $T=300$ K in Fe, Ni, Co, FePt and FePd for
different magnetization directions.  The bar diagrams in the first row of (a),
(b), (d)-(f) show the values of $\alpha=\alphaic+\alphasj$ 
at different energy positions.  The true Fermi energy
level in each material lies at the position of the middle vertical line.  The
line diagrams in the lower row of (a), (b), (d)-(f) depict the energy
dependence of $\sigma=\sigmaic+\sigmasj$.  In (b),
hatched bars and dotted lines stand for GGA$+U$ calculations in Ni with $U=3.9$
eV and $J=1.1$ eV.  In (d)-(f), hatched bars and dotted lines stand for
calculations in Co, FePt and FePd with the magnetization pointing along the
[110] direction.  A sketch of the experimental setup is shown in (c).}
\label{fig:edep}
\end{figure*}

Our approach is based on electronic structure calculations performed within the
full-pot\-en\-tial linearized augmented plane-wave method as implemented in the
J\"ulich density functional theory code {\tt FLEUR}.\cite{fleur}  The matrix
elements of the multi-band Bloch Ham\-il\-ton\-i\-an $\hat H(\kf)$ in the basis
of max\-i\-mal\-ly-loc\-al\-iz\-ed Wannier
functions\cite{Freimuth:2008,Mostofi:2008} have been computed using the Wannier
interpolation technique\cite{Yates:2007} and inserted into the equations for
$\hat{\sigma}^{\q{ic}}$ and $\hat{\sigma}^{\q{sj}}$ obtained within the Kubo-St\v{r}eda
formalism,\cite{Streda:1975} assuming short-range disorder in the
system.\cite{Sinova:2010}  The knowledge of any free parameters is not required
in this scheme. For the evaluation of the integral in \eq{eq:Mott}, we adopted
an energy grid that was denser at low temperatures, since the energy derivative
of the Fermi function becomes a $\delta$-distribution in this limit.  Near room
temperature, we found that an energy spacing of $\Delta\vareps\approx 5$ meV
offered the best trade-off between accuracy and computational cost, leading
to an error of about 2\% for $\hat\alpha$.

It follows from an argument by Berger\cite{Berger:1972} that large values 
for the components of
$\hat\alpha$ could not arise if electrons of different energies experienced an
AHE of the same magnitude and sign, since the transverse velocities of electrons
diffusing down the applied temperature gradient would then cancel with those of
the less energetic electrons diffusing up the temperature gradient and the net
transverse current would be zero.  The same can also be deduced from the Mott
formula \eq{eq:Mott}: Since the product $(\partial_{\mu}f)\cdot(\vareps-\mu)$
is an antisymmetric function with respect to the Fermi energy level
$\vareps_{F}=\mu$, the ANE would vanish if the AHE conductivity
$\sigma^{\q{AHE}}$ was a symmetric function around $\vareps_{F}$, i.e., if
it was equal for a pair of energy values with the same distance to
$\vareps_{F}$.

In Fig.~\ref{fig:edep}(a), (b), (d)-(f), the component 
$\sigma\equiv\mf\sigma\cdot\hat{\mf M}$ 
of the anomalous Hall vector $\mf\sigma$
and the component 
$\alpha\equiv\mf\alpha\cdot\hat{\mf M}$  of the thermoelectric conductivity vector $\mf\alpha$ 
parallel to the magnetization direction $\hat{\mf M}\equiv\mf M/|\mf M|$
are shown. They have been computed from \eq{eq:Mott} at room temperature
as a function of the Fermi energy level
in Fe, Ni, Co, FePt and FePd.
For the [001] magnetization direction, 
$\sigma$ and $\alpha$ correspond to the tensor elements 
$\sigma_{xy}$ and $\alpha_{xy}$, while for the [110] magnetization
direction, they correspond to the tensor elements 
$(\sigma_{yz}+\sigma_{zx})/\sqrt{2}$ and
$(\alpha_{yz}+\alpha_{zx})/\sqrt{2}$.\cite{supplement}
In bcc Fe, we observe that the intrinsic contribution $\sigmaic$ is nearly
symmetric around the Fermi level $\vareps_{F}$, resulting in a
rather small value of $\alphaic$. If we shift the Fermi energy
artificially by $+$0.08 eV, the course of $\sigmaic$ gets more asymmetric
with respect to the new Fermi energy and the value of $\alphaic$ increases
significantly by a factor of 6.  As a consequence, the total value
$\alphaic+\alphasj$ changes sign and becomes positive. If the Fermi
energy level is lowered by $-$0.12 eV instead, all contributions become negative.
In general, we find that the thermoelectric conductivity in Fe, Co and Ni
appears to be highly sensitive to the position of the Fermi level, suggesting
that the ANE in ferromagnets can be easily tuned by~e.g.~suitable doping.  On
the other hand, for the compounds FePd and FePt the Fermi energy dependence of
$\alpha$ is much less pronounced.

\begin{table}[b]
\caption{Comparison of $Q_{1}^{\q{ic+sj}}$ 
with experimental data for $Q_{1}$ 
in units of $10^{-11}$V/(K$\cdot$G)
near room temperature. For Ni, the values of $U$ and $J$ are in eV.}
\label{tab:exp}
\begin{ruledtabular}
\begin{tabular}{lrrr|r}
  & $Q_{1}^{\q{ic}}$
  & $Q_{1}^{\q{sj}}$ 
  & $Q_{1}^{\q{ic+sj}}$
  & $Q_{1}$ Expt.\cite{supplement}
   \\ \hline
Fe                  &  $-$0.52 & $-$0.39 & $-$0.91 & $-$0.81 to $-$2.08 \\ \hline
Co   [001]          &     1.00 &    0.07 &    1.07 &    2.00 to \phantom{$-$}2.19 \\
\phantom{Co} [110]  &     0.24 &    0.03 &    0.27 &            \\\hline
Ni ${U=0.0},\,{J=0.0}$           & $-$15.12 & $-$1.09 & $-$16.20 & 3.04 to \phantom{$-$}7.31 \\
\phantom{Ni} ${U=1.9},\,{J=1.1}$ & $-$4.40 &     1.98 & $-$2.42  & \\
\phantom{Ni} ${U=3.9},\,{J=1.1}$ & $-$0.28 &     1.48 &    1.20  & \\
\phantom{Ni} ${U=3.9},\,{J=2.6}$ &    0.74 &     1.90 &    2.64  & \\\hline
FePt                             &    4.61 &     1.27 &    5.88  &  5.60 \\ 
\end{tabular}
\end{ruledtabular}
\end{table}

By now it is established that the side-jump contribution to the AHE is
important in FePd whereas the intrinsic AHE is dominant in
FePt.\cite{He:2012,Seemann:2010} As follows from Fig.~\ref{fig:edep}(e)-(f),
this statement also applies to the ANE in these materials, i.e.,
$\alphasj(\vareps_{F})$ is as large as $\alphaic(\vareps_{F})$ in
FePd, but only half of this value in FePt. This crossover behavior is caused by
the different spin-orbit interaction strength of Pd and Pt
atoms.\cite{Seemann:2010,Zhang:2011a} In Fe and Co, the magnitude of
$\alphasj(\vareps_{F})$ is greater than that of
$\alphaic(\vareps_{F})$, albeit $\sigmasj(\vareps_{F})$ being smaller
than $\sigmaic(\vareps_{F})$ in both materials.

In analogy to the AHE,\cite{Zhang:2011} one might suspect that the ANE is
highly anisotropic with respect to the direction of the magnetization in the
crystal for hcp Co, and $L1_{0}$ ordered FePd and FePt alloys, due to their
uniaxial crystal structure. Indeed, at the Fermi energy, the side-jump
contribution $\alphasj$ switches its sign in Co and is distinctly reduced in
FePd and FePt as the direction of the magnetization is changed from [001] to
[110] direction.  However, the anisotropy of the intrinsic contribution
$\alphaic$ is not that strong.  Such a different dependence of
$\alphaic$ and $\alphasj$ on the magnetization direction may be
attributed to the different distribution of $\sigmaic$ and $\sigmasj$
in the Brillouin zone of these materials.\cite{Weischenberg:2011}

The ANE in fcc Ni presents an exceptional case, since the theoretical value of
the thermoelectric conductivity in this material is much larger than in other
considered compounds, see Fig.~\ref{fig:edep}(b). In Ni, the intrinsic
anomalous Hall conductivity is sharply peaked near the Fermi energy, and the
respective value for the ANE depends on which side of the peak it is evaluated.
There are many indications, however, that the large value of $\sigmaic$ in
Ni is mainly an artifact of the local density approximation (LDA) or the
generalized gradient approximation (GGA),  because correlation effects among
the 3$d$ electrons in this material become of crucial importance for its
properties.\cite{Yang:2001,Fuh:2011,Weischenberg:2011} For this reason, we have
adopted the same approach as in our previous work and took the correlation
effects into account within the GGA$+U$ scheme.\cite{Shick:1999}  For the
intra-atomic Coulomb repulsion and exchange parameters $U$ and $J$, we chose 
values up to 3.9~eV and 1.1~eV, respectively.  This choice of parameters has been
found to greatly improve the agreement of the calculated AHE in Ni.
Fig.~\ref{fig:edep}(b) reveals that correlations have also a significant effect on
the energy dependence of $\sigma^{\q{AHE}}$ and $\alpha$ in Ni.  We
observe that the peak in the intrinsic contribution to the AHE flattens out,
whereas the side-jump contribution remains mostly unaffected upon including the
$U$. This can again be understood from the different behavior of the two
effects at the Fermi surface.\cite{Weischenberg:2011}  Upon including $U$, the
intrinsic contribution $\alphaic$ and the side-jump contribution
$\alphasj$ change their sign and the magnitude of the
thermoelectric conductivity is greatly reduced.

For comparison with experiment, we consider the situation in which a
temperature gradient in $\hat x$ direction, $-\partial_{x}T$, is applied to
an electrically isolated sample perpendicular to the magnetization
$\mf M\parallel\hat{z}$, see Fig.~\ref{fig:edep}(c).  As a function of the magnetic
field strength $|\mf H|=H_z$ and magnetization $|\mf M|=M_z$, 
the Nernst effect obeys a law of the type
$E_{y}/(-\partial_{x}T)=H_{z}Q_{0}+4\pi M_{z}Q_{1}$, where $Q_{0}$ and $Q_{1}$
are the ordinary and anomalous Nernst coefficients,
respectively.\cite{Jan:1957} However, in ferromagnetic materials, the ordinary
Nernst coefficient is very small,\cite{Nielsen:1934} $Q_{0}\ll Q_{1}$.  The
remaining coefficient $Q_{1}$ is generally measured in a zero-current
configuration, $j_{x}=j_{y}=0$, with the boundary condition
$\partial_{y}T=0$.\cite{Butler:1940}  For a spatially uniform sample in an
experimental setup as depicted in Fig.~\ref{fig:edep}(c), it holds that
$\sigma_{xx}=\sigma_{yy}$, $\sigma_{xy}=-\sigma_{yx}$ and likewise for the
components of $\hat\alpha$. In this scenario, we obtain from \eq{eq:j}  
\begin{align}
4\pi M_{z}Q_{1}&=\rho_{xx}(\alpha_{xy}-S\sigma_{xy})
\label{eq:Q}\\
&=\rho_{xx}(\alpha^{\q{ic}}_{xy}-S\sigma^{\q{ic}}_{xy})
 +\rho_{xx}(\alpha^{\q{sj}}_{xy}-S\sigma^{\q{sj}}_{xy}),\nonumber
\end{align}
where $\rho_{xx}=1/\sigma_{xx}$ is the resistivity of the sample and the
so-called See\-beck-co\-ef\-fi\-ci\-ent is defined by $S\equiv
E_{x}/\partial_{x}T=\alpha_{xx}/\sigma_{xx}$.  The last line of \eq{eq:Q} may
be interpreted as $4\pi M_{z}(Q_{1}^{\q{ic}}+Q_{1}^{\q{sj}})$, where the
intrinsic contribution to the anomalous Nernst coefficient is denoted by
$Q_{1}^{\q{ic}}$ and the side-jump contribution is denoted by $Q_{1}^{\q{sj}}$.
While the Seebeck-co\-ef\-fi\-ci\-ent $S$ describes the conversion of a
thermal current into a longitudinal electrical current, the
Nernst-co\-ef\-fi\-ci\-ent $Q_{1}$ is a measure of the corresponding
transverse effect. 
Even though the value of the temperature gradient $-\partial_{x}T$ 
and the strength of
the magnetic induction $B_{z}$ do not appear in \eq{eq:Q}, they seem to have 
a great influence on the ANE experimentally.\cite{Campbell:1908,Campbell:1923}
In particular, $B_{z}$ influences the magnitude of the magnetization
and the corresponding electronic structure.
However, values for $Q_{1}$, $\rho_{xx}$, $\alpha_{xy}$, $S$, $\sigma_{xy}$ and $M_{z}$
were not measured simultaneously in most experiments.  
We therefore gathered values for the resistivity $\rho_{xx}$ and Seebeck
coefficient $S$ from various sources and computed the scattering-independent
contribution $Q_{1}^{\q{ic+sj}}=Q_{1}^{\q{ic}}+Q_{1}^{\q{sj}}$ 
to the anomalous Nernst coefficient
following \eq{eq:Q} (Details of this procedure can be found
in the supplement).\cite{supplement}

A comparison of our calculated values with experimental data is presented in
\tab{tab:exp}.  The experimental values show a considerable spread, which
reflects the fact that the ANE is found to depend sensitively on experimental
details and material-specific parameters.  Nevertheless, it can clearly be seen
that the inclusion of the side-jump contribution is crucial for Fe and FePt and
lets theory and experiment match very well: $Q_{1}^{\q{ic+sj}}$ is about 112\%
of the smaller experimental value in Fe and 105\% of the experimental value in
FePt.  For Co, the side-jump contribution to $Q_{1}^{\q{ic+sj}}$ is relatively
small and only around 10\% for both magnetization directions, but it still
brings the theoretical value for the anomalous Nernst coefficient closer to
experiment.  The lack of better agreement may be due to the fact that for the
Seebeck coefficient in Co
solely experimental data for polycrystalline samples has been available,
but the corresponding value for monocrystalline samples
should be inserted into \eq{eq:Q} instead. 
Remarkably, when the magnetization is changed from [001] into [110] direction,
the sign change of $\alphasj$ in Co at the Fermi energy level is
compensated by the sign change of $\sigmasj$, and the net contribution
$Q_{1}^{\q{sj}}$ stays roughly the same.  For Ni, the value calculated in bare GGA
differs drastically from experiment, and even has the wrong sign. However, when
the value of $U$ is increased within GGA$+U$, the calculated value approaches
the experimental result in magnitude and sign. This
suggests that the main reason for the discrepancy between experiment and theory
in Ni originates from an inadequate description of the electronic structure
in the vicinity of the Fermi energy level within GGA.\cite{footnote}

\begin{figure}
\includegraphics[width=0.48\textwidth]{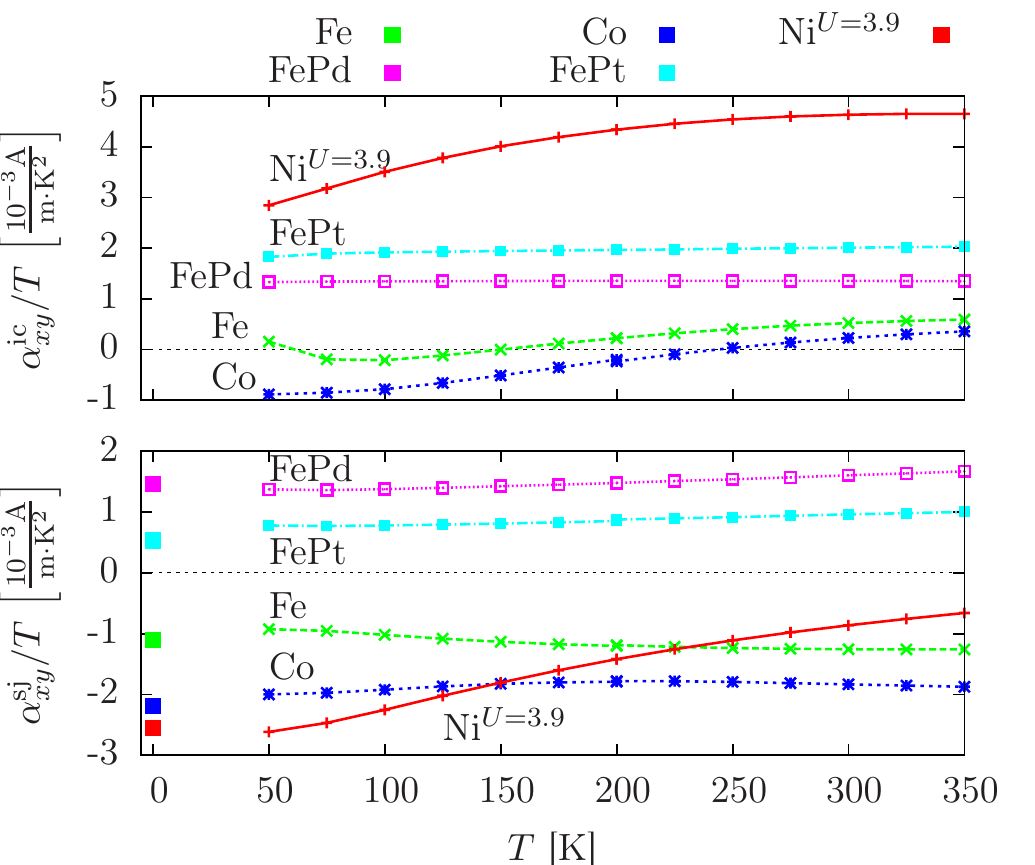}
\caption{(Color online) Temperature dependence of the in\-trin\-sic- and side-jump
contribution to the ANE. The squares at $T=0$ K stand for 
the values of 
$\alpha^{\q{sj}}_{xy}/T$ computed from \eq{eq:zerotemp}.}
\label{fig:tdep}
\end{figure}

Overall, the values in \tab{tab:exp} demonstrate that the intrinsic- and
side-jump contribution play an important role in the ANE of ferromagnets.  This
finding is consistent with earlier studies in this field, which examined the
behavior of the anomalous Nernst coefficient $Q_{1}$ as a function of the
resistivity $\rho_{xx}$. As far as the  scattering-independent contributions
are concerned, one would expect a linear dependence of the
form $Q_{1}/T\propto \rho_{xx}$,\cite{Berger:1972} which is also observed in
experiment.\cite{Campbell:1979,Lucinski:1993,Kondorskii:1964,Cheremushkina:1966,
Vasileva:1972} Our work substantiates these observations with quantitative
analysis.

The calculated temperature dependence of the thermoelectric conductivity tensor
for [001] magnetization direction is depicted in \fig{fig:tdep}.
$\alpha^{\q{ic}}_{xy}$ is positive at 300 K, but
changes its sign in Fe and Co as the temperature is decreased. Below 50 K, it
becomes positive again in Fe. The importance of the side-jump contribution to
the ANE is stressed by the fact that for considered materials  $\alpha^{\q{sj}}_{xy}$
is of the same order of magnitude or even larger than $\alpha^{\q{ic}}_{xy}$.  In
contrast to the intrinsic contribution, $\alpha^{\q{sj}}_{xy}$ does not change its
sign. While the temperature dependence of $\alpha^{\q{sj}}_{xy}/T$ 
is almost absent in FePd and FePt, it ranges from 
$-2.6\cdot10^{-3}$A/(m$\cdot$K$^2$) at $T=50$~K to
$-0.7\cdot10^{-3}$A/(m$\cdot$K$^2$) at $T=350$~K in Ni.

In the zero temperature limit, one can apply the Sommerfeld expansion to the
integral in \eq{eq:Mott} to obtain the standard Mott formula which relates the
ANE to the energy derivative of the AHE:\cite{Ashcroft:1988}
\bege
\frac{\alpha_{xy}}{T}=-
\frac{\pi^{2}k_{B}^{2}}{3e}
\bracks{\frac{d\sigma_{xy}}{d\vareps}}_{\vareps=\vareps_{F}}.
\label{eq:zerotemp}
\ee
For the intrinsic Nernst effect, it is known that the above formula may be
violated as $T\rightarrow 0$ K.\cite{Bergman:2010}  Indeed, we find that the
energy derivative of the intrinsic contribution converges only very slowly with
respect to the number of $\kf$-points in the Brillouin zone that are used for
the evaluation of $\sigma^{\q{ic}}$.  The slow convergence is due to the
sensitivity of  the Berry curvature to the position of the Fermi energy,
especially when the latter approaches avoided band crossings or points of band
degeneracy.\cite{Yao:2004}  However, for the side-jump contribution, Mott's
formula \eq{eq:zerotemp} holds, as can be seen by interpolating the curves in
\fig{fig:tdep} to $T=0$ K. Apart from Ni, it yields a rather good estimate for
the value of $\alpha^{\q{sj}}_{xy}/T$ at room temperature as well.

In summary, we presented the {\it ab initio}  calculations of the scattering
independent contributions to the ANE in several ferromagnets.  The theoretical
values for the thermoelectric conductivity tensor and the comparison of the
calculated anomalous Nernst coefficient with experiment suggests that the ANE
in elementary Fe, Co, Ni, in the ferromagnetic alloy FePt and presumably also
in FePd is largely caused by the intrinsic and side-jump mechanisms.
Discrepancies between theory and experiment in Ni are likely due to the
imprecise description of correlation effects within bare GGA, which can be
remedied by GGA$+U$ calculations.

We thank J. Sinova for fruitful discussions and
gratefully acknowledge J\"ulich Supercomputing Centre for computing time as 
well as funding by the HGF-YIG programme VH-NG-513. J.~W. was supported under 
grant SPP
1538 SpinCaT by the German Science Foundation.


\begin{thebibliography}{47}
\bibitem [{foo()}]{footnote}%
  \BibitemOpen
  \href@noop {} {}\bibinfo {howpublished} {Further reasons for the discrepancy
  might be the temperature dependence of the electronic structure or details of
  dynamical Coulomb screening of $d$-electrons in Ni, etc.}\BibitemShut {Stop}%
\end{thebibliography}
\end{document}


\title{Supplement}

\author{J\"urgen Weischenberg}
\email[corresp.~author:~]{j.weischenberg@fz-juelich.de}
\affiliation{Peter Gr\"unberg Institut \& Institute for Advanced Simulation,
Forschungszentrum J\"ulich and JARA, 52425 J\"ulich, Germany}

\author{Frank Freimuth}
\affiliation{Peter Gr\"unberg Institut \& Institute for Advanced Simulation,
Forschungszentrum J\"ulich and JARA, 52425 J\"ulich, Germany}

\author{Stefan Bl\"ugel}
\affiliation{Peter Gr\"unberg Institut \& Institute for Advanced Simulation,
Forschungszentrum J\"ulich and JARA, 52425 J\"ulich, Germany}

\author{Yuriy Mokrousov}
\affiliation{Peter Gr\"unberg Institut \& Institute for Advanced Simulation,
Forschungszentrum J\"ulich and JARA, 52425 J\"ulich, Germany}

\date{\today}

\maketitle

We compare the theoretical value of the anomalous Nernst coefficient $Q_1$ with
experimental data.  For this purpose, we need the values of the resistivity
$\hat\rho$, the thermoelectric conductivity tensor $\hat\alpha$, the Seebeck
coefficient $S$, the conductivity $\hat\sigma$ and the magnetization $\mf M$.  In
linear transport theory, the temperature dependent transport coefficients
$\sigma_{ij}=e\mathcal L^{(0)}_{ij}$ and $\alpha_{ij}=
\mathcal L^{(1)}_{ij}/T$ can be computed from
the formula
\bege
\label{eq:transport}
\mathcal L^{(n)}_{ij}(T)=-\frac{1}{e}\int d\vareps\,
\frac{\partial f(\vareps,\mu,T)}{\partial\mu}
\cdot\sigma^{T=0}_{ij}(\vareps)\cdot\parens{\vareps-\mu}^{n},
\ee
where the expression for the Fermi distribution function is given by
$f(\vareps,\mu,T)=1/[\exp((\vareps-\mu)/k_{B}T)+1]$.  

Since the derivate $\partial f/\partial\mu$ decays exponentially, the integrand
in \eq{eq:transport} needs only be evaluated in an interval of about $\pm 0.5$
eV around the Fermi energy level.  For $\sigma^{T=0}_{xy}$ we insert the anomalous
Hall conductivity $\sigma^{\q{AHE}}=\sigmaic+\sigmasj$, which can be computed
from the electronic structure of the pristine crystal alone with the same
method as developed in Ref.~(\onlinecite{Weischenberg:2011}).  In our
electronic structure calculations we have chosen a plane-wave cutoff
$K_{\q{max}}$ of 3.8 bohr$^{-1}$ and constructed a set of 18 maximally
localised Wannier functions per atom on an uniform $8\times 8 \times 8$ grid of
{\it ab initio} $\kf$-points using the {\tt WANNIER90} code and our interface
between {\tt FLEUR} and {\tt WANNIER90}.\cite{Mostofi:2008,Freimuth:2008} The
actual calculation of $\sigma^{\q{AHE}}$ has then been performed on a very
dense grid of up to $800\times 800\times 800$ $\kf$-points using the Wannier
interpolation scheme.\cite{Yates:2007}


For $|\mf M|$ we take the saturation magnetization at room temperature.  Since
the Curie temperature $T_{C}$ for Fe, Co and Ni is quite large, $T_{C}\gg 300$
K, we suppose that the temperature dependence of $|\mf M|$ has no significant
effect on the calculated values for $Q_{1}^{\q{ic+sj}}$. 
For a general magnetization along $\hat{\mf M}\equiv\mf M/|\mf M|$, we introduce the 
anomalous Hall vector $\mf \sigma$ and the thermoelectric conductivity vector $\mf \alpha$,
\bege{}
[\mf\sigma]_i\equiv\frac{1}{2}\epsilon_{ijk}\sigma_{jk},\quad 
\mf\sigma=\parens{\begin{matrix}\sigma_{yz}\\\sigma_{zx}\\\sigma_{xy}\end{matrix}},\qquad
[\mf\alpha]_i\equiv\frac{1}{2}\epsilon_{ijk}\alpha_{jk},\quad 
\mf\alpha=\parens{\begin{matrix}\alpha_{yz}\\\alpha_{zx}\\\alpha_{xy}\end{matrix}},
\ee
where $\epsilon_{ijk}$ are the components of the Levi-Civita tensor. 
These vectors can be decomposed into
a part which is parallel and into a part which is perpendicular to $\hat{\mf M}$,
\bege
\mf \sigma =\sigma_{\parallel}\hat{\mf M}+\sigma_\perp\hat{\mf n},\quad
\mf \alpha =\alpha_{\parallel}\hat{\mf M}+\alpha_\perp\hat{\mf n}.
\ee
In the above equation, $\hat{\mf n}$ denotes the unit vector perpendicular to
$\hat{\mf M}$.  In single crystals, $\mf \sigma$, $\mf \alpha$ and $\mf M$ are
perfectly colinear only when $\mf M$ points into special high-symmetry
directions.  For the [001] magnetization direction, it has been shown in the
main text that the anomalous Nernst coefficient can be computed via the
equation
\bege
Q_{1}=\rho_{xx}(\alpha_{xy}-S\sigma_{xy})/4\pi M_{z}.
\ee
For other magnetization directions, we replace $\hat\sigma_{xy}$ with the parallel component
$\sigma=\mf\sigma\cdot\hat{\mf M}$ and $\hat\alpha_{xy}$ with the parallel component
$\alpha=\mf\alpha\cdot\hat{\mf M}$. For example, if the magnetization points into
[110] direction, these correspond to the tensor elements
$\sigma=(\sigma_{yz}+\sigma_{zx})/\sqrt{2}$ and
$\alpha=(\alpha_{yz}+\alpha_{zx})/\sqrt{2}$, respectively.
The anisotropy of $\hat\rho$ and $S$ needs also be considered. In hcp Co, in the
presence of a magnetic field, the resistivity in the direction of the $c$-axis
is about 10.32~$\mu\Omega\cdot$cm, but drops to 5.55~$\mu\Omega\cdot$cm in the
direction perpendicular to this axis.\cite{Masumoto:1966}  For the anisotropy
of the Seebeck coefficient in hcp Co, no suitable experimental values have been
found.  We have therefore resorted to data for polycrystalline samples.
Presumably, the anisotropy of $S$ in hcp Co is relatively
small.\cite{Sinha:1991}

\begin{figure}[t]
\includegraphics[width=0.32\textwidth]{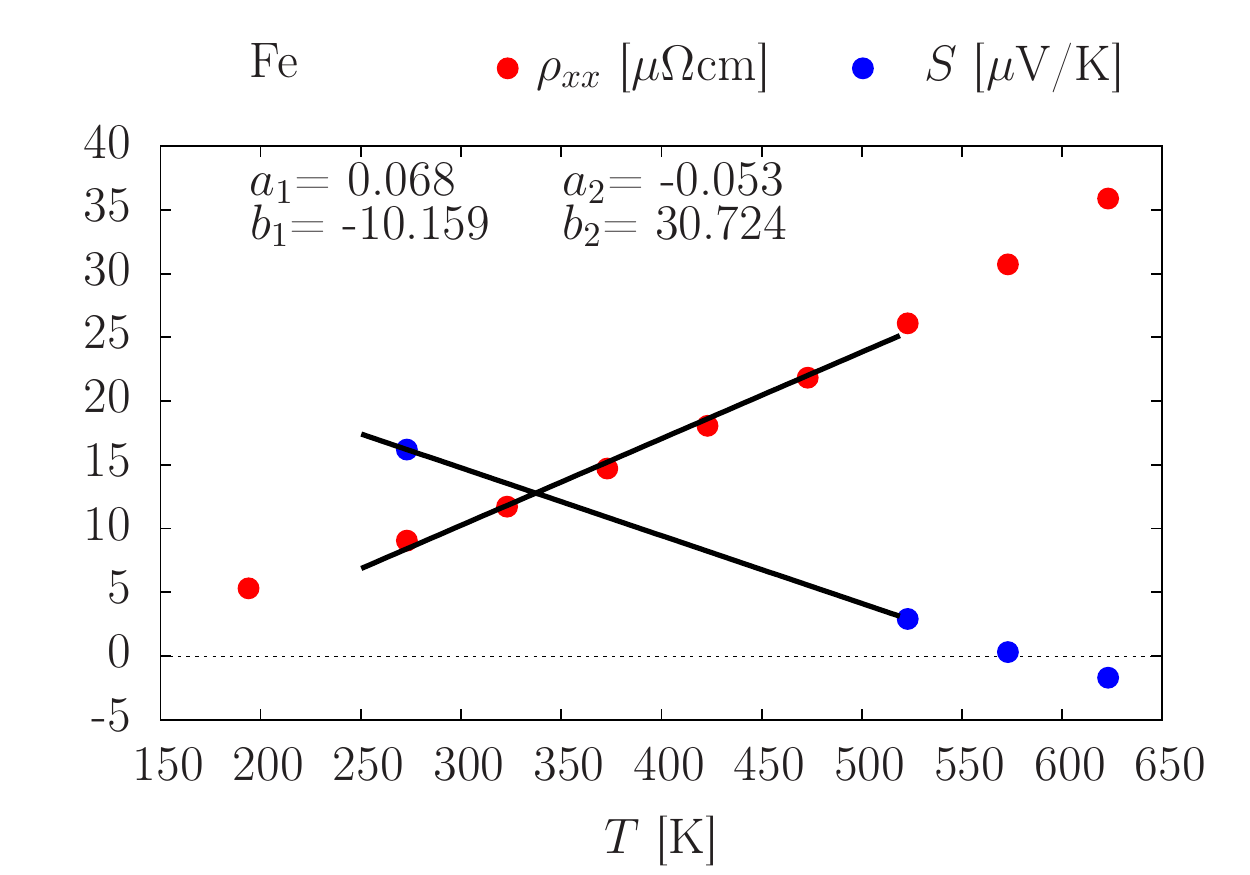}
\includegraphics[width=0.32\textwidth]{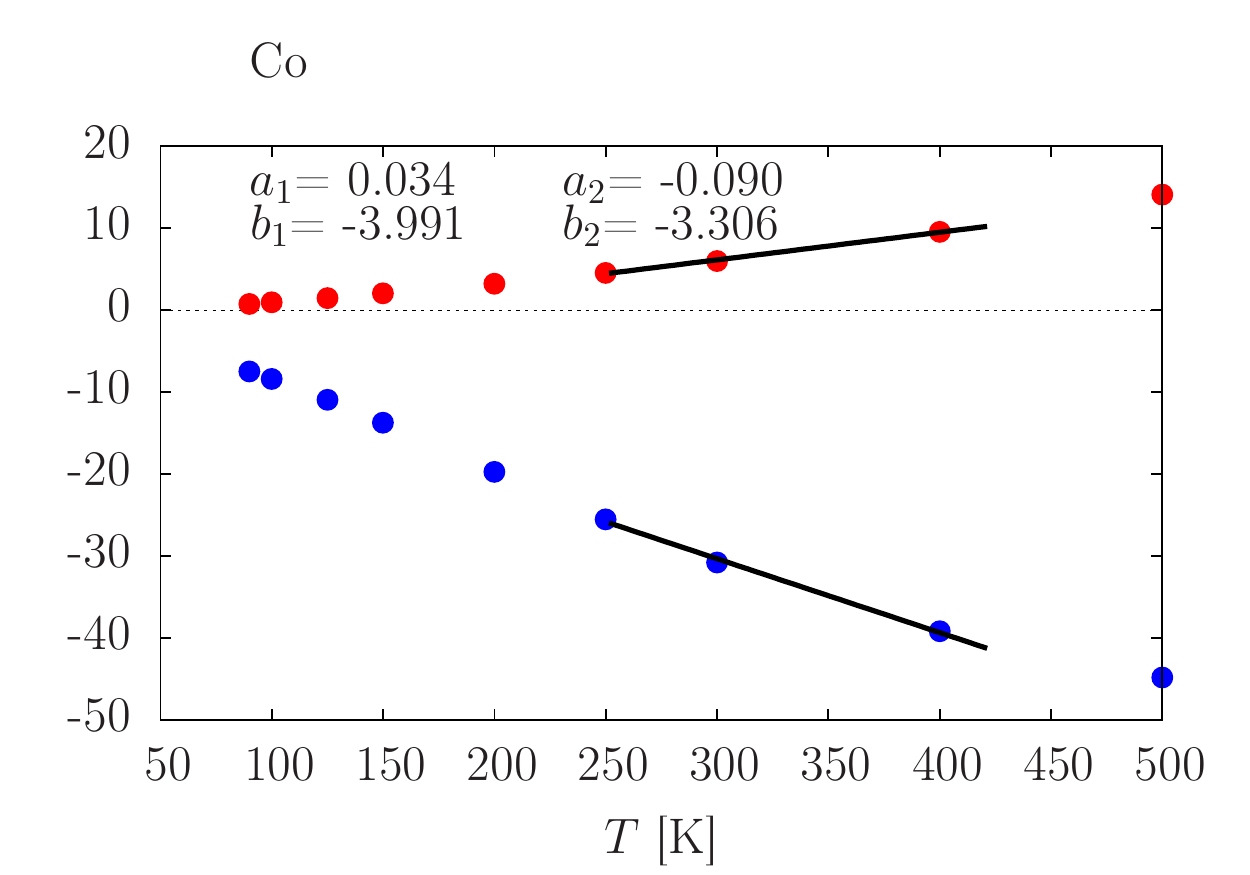}
\includegraphics[width=0.32\textwidth]{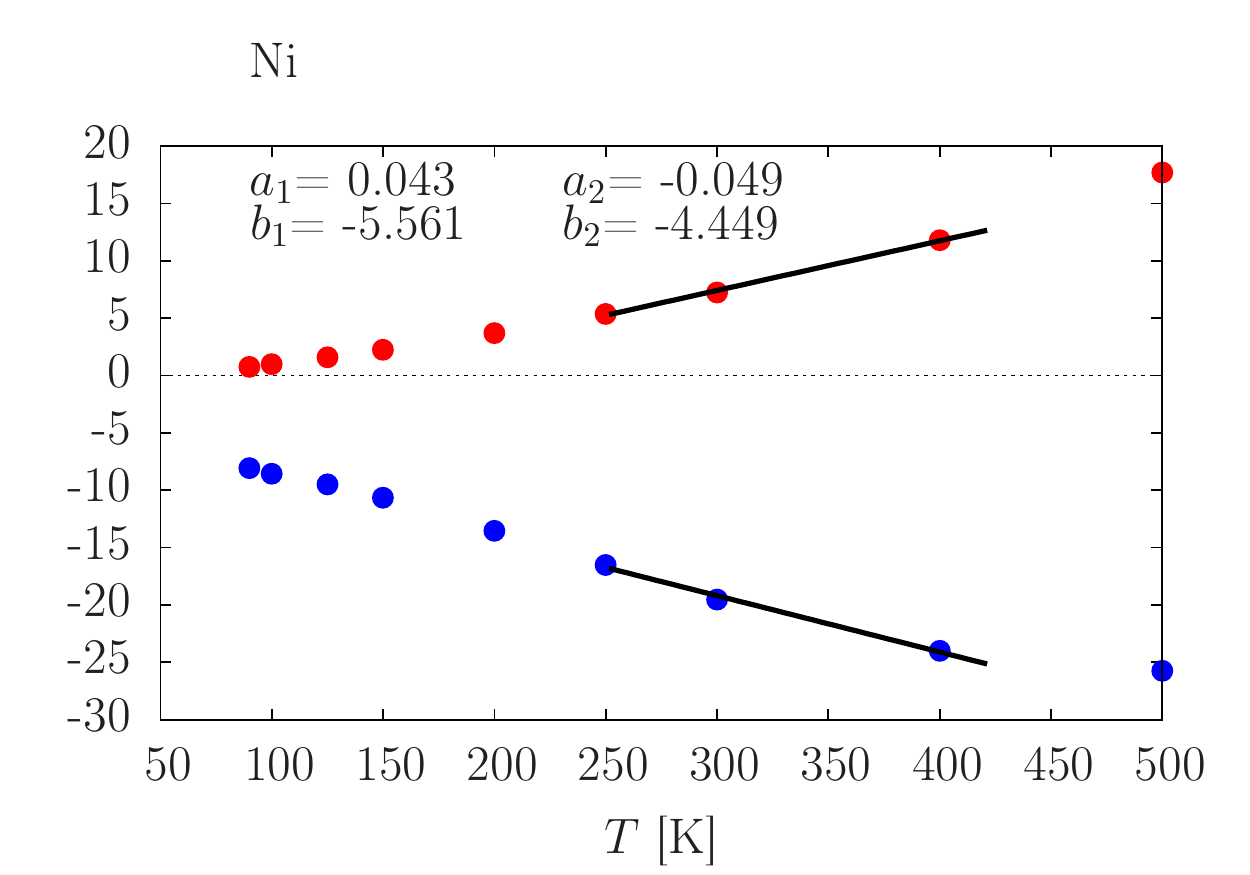}
\caption{Interpolation of $\rho_{xx}$ and $S$ for Fe, Co and Ni 
based on data from Ref.~(\onlinecite{Fulkerson:1966,Laubitz:1973,Laubitz:1976})
for polycrystalline samples.}
\label{fig:interpolation}
\end{figure}

In most experiments, the values of $Q_{1}$, $\hat\rho$, $S$ and $\mf M$ are not
measured simultaneously. As illustrated in \fig{fig:interpolation}, we
therefore estimate $\rho_{xx}$ and $S$ from linear interpolation whenever their
values are not available:
\bege
\rho_{xx}&=a_{1}\cdot T+b_{1}, & S&=a_{2}\cdot T+b_{2}.
\ee
The uncertainities in $\hat\rho$ and $S$ also affect our prediction of the
theoretical values for $Q_{1}^{\q{ic+sj}}$. However, the spread of the
experimental values for $Q_1$ appears to be much more significant. A comparison
of theoretical with experimental data can be found in \tab{tab:Exp} and
\tab{tab:Exp2}.

\newcommand{\cB}[1]{{\color{NavyBlue} #1 }}
\newcommand{\cG}[1]{{\color{ForestGreen} #1 }}
\newcommand{\cR}[1]{{\color{BrickRed} #1 }}
\newcommand{\as}{$^*$}

\begin{table}[b]
\caption{Comparison with experimental data. 
Temperature $T$ in K; 
magnetic field $B$ in G; 
longitudinal resistivity $\rho$ in $\mu\Omega\cdot$cm;
Seebeck coefficient $S$ in $\mu$V/K;
transverse anomalous Hall conductivity $\sigma$ in $1/(\Omega\cdot$cm);
transverse anomalous thermoelectric conductivity $\alpha$ in 
$10^{-3}$A/(m$\cdot$K);
anomalous Nernst coefficient $Q_1$ in $10^{-11}$V/(K$\cdot$G).
Values that are labeled by an asterix are taken from interpolation of data from
Ref.~(\onlinecite{Fulkerson:1966,Laubitz:1973,Laubitz:1976,Masumoto:1966}).  The sign of the values
from Ref.~(\onlinecite{Zahn:1904}) has been adjusted to be in accordance with the sign
convention of Campbell.\cite{Campbell:1923}}
\label{tab:Exp}
\begin{tabular*}{\textwidth}{@{\extracolsep{\fill}}lrrdd|rrrd>{\cG\bgroup}r<{\egroup}>{\cB\bgroup}r<{\egroup}>{\cR\bgroup}r<{\egroup}|d}
\hline\hline
  & \multicolumn{1}{c}{$T$}
  & \multicolumn{1}{c}{$B$}
  & \multicolumn{1}{c}{$\rho$} 
  & \multicolumn{1}{c|}{$S$} 
  & \multicolumn{1}{c}{$\sigmaic$} 
  & \multicolumn{1}{c}{$\alphaic/T$} 
  & \multicolumn{1}{c}{$\sigmasj$} 
  & \multicolumn{1}{c}{$\alphasj/T$} 
  & \multicolumn{1}{c}{\cG{$Q_{1}^{\q{ic}}$}}
  & \multicolumn{1}{c}{\cB{$Q_{1}^{\q{sj}}$}}
  & \multicolumn{1}{c|}{\cR{$Q_{1}^{\q{ic+sj}}$}}    
  & \multicolumn{1}{c}{$Q_{1}$ Expt.}    
   \\ \hline
Fe          & 291&  6290&  9.63^*&  15.24^* &  744 & 0.505  &   115 &  -1.254 & $-$0.43  & $-$0.24  & $-$0.67&  -1.05  \cite{Zahn:1904}  \\   
            & 293& ---  & 12.55  &  14.29   &  744 & 0.509  &   115 &  -1.254 & $-$0.52  & $-$0.30  & $-$0.82&  -0.82  \cite{Hall:1911}  \\
            & 313& 23000& 11.65  &  14.07^* &  741 & 0.544  &   115 &  -1.258 & $-$0.46  & $-$0.29  & $-$0.75&  -2.04  \cite{Butler:1940}\\
            & 313&  --- & 13.73  &  13.70   &  741 & 0.544  &   115 &  -1.258 & $-$0.52  & $-$0.34  & $-$0.87&  -0.90  \cite{Hall:1911}  \\
            & 323& 23000& 12.21  &  13.54^* &  739 & 0.559  &   116 &  -1.260 & $-$0.45  & $-$0.31  & $-$0.76&  -2.06  \cite{Butler:1940}\\
            & 333& 18000& 12.49^*&  13.01^* &  738 & 0.572  &   116 &  -1.260 & $-$0.43  & $-$0.32  & $-$0.76&  -0.81  \cite{Smith:1916} \\
            & 333& 23000& 12.77  &  13.01^* &  738 & 0.572  &   116 &  -1.260 & $-$0.44  & $-$0.33  & $-$0.77&  -2.08  \cite{Butler:1940}\\
            & 333&  --- & 14.91  &  13.07   &  738 & 0.572  &   116 &  -1.260 & $-$0.52  & $-$0.39  & $-$0.91&  -0.98  \cite{Hall:1911}  \\
            & 343& 23000& 13.33  &  12.48^* &  737 & 0.583  &   116 &  -1.261 & $-$0.43  & $-$0.35  & $-$0.78&  -2.10  \cite{Butler:1940}\\\hline
Co [001]    & 297&  6290& 10.32^*& -30.09^* &  484 & 0.216  &   222 &  -1.831 &    0.88  &    0.07  &    0.95&   2.00  \cite{Zahn:1904}  \\
            & 297& 10570& 10.32^*& -30.09^* &  484 & 0.216  &   222 &  -1.831 &    0.88  &    0.07  &    0.95&   1.80  \cite{Zahn:1904}  \\
            & 298&  9000& 10.32^*& -30.18^* &  484 & 0.220  &   222 &  -1.832 &    0.88  &    0.07  &    0.95&   1.90  \cite{Hall:1925}  \\
            & 300&  --- & 10.32^*& -30.36^* &  484 & 0.226  &   222 &  -1.833 &    0.89  &    0.07  &    0.96&   1.80  \cite{Smith:1911} \\
            & 320&  9000& 10.32^*& -32.16^* &  485 & 0.286  &   223 &  -1.850 &    0.95  &    0.07  &    1.03&   2.19  \cite{Hall:1925}  \\
            & 333& 18000& 10.32^*& -33.33^* &  486 & 0.320  &   223 &  -1.861 &    1.00  &    0.07  &    1.07&   2.00  \cite{Smith:1916} \\
            & 350&  --- & 10.32^*& -34.87^* &  486 & 0.359  &   223 &  -1.874 &    1.05  &    0.07  &    1.12&   2.15  \cite{Smith:1911} \\\hline
Co [110]    & 297&  6290&  5.55^*& -30.09^* &  125 & 1.010  & $-$91 &   1.657 &    0.21  &    0.07  &    0.28&   2.00  \cite{Zahn:1904}  \\
            & 297& 10570&  5.55^*& -30.09^* &  125 & 1.010  & $-$91 &   1.657 &    0.21  &    0.07  &    0.28&   1.80  \cite{Zahn:1904}  \\
            & 298&  9000&  5.55^*& -30.18^* &  125 & 1.009  & $-$91 &   1.644 &    0.21  &    0.07  &    0.28&   1.90  \cite{Hall:1925}  \\
            & 300&      &  5.55^*& -30.36^* &  126 & 1.007  & $-$90 &   1.618 &    0.21  &    0.07  &    0.28&   1.80  \cite{Smith:1911} \\
            & 320&  9000&  5.55^*& -32.16^* &  128 & 0.994  & $-$88 &   1.359 &    0.23  &    0.05  &    0.27&   2.19  \cite{Hall:1925}  \\
            & 333& 18000&  5.55^*& -33.33^* &  130 & 0.995  & $-$87 &   1.197 &    0.24  &    0.03  &    0.27&   2.00  \cite{Smith:1916} \\
            & 350&      &  5.55^*& -34.87^* &  133 & 1.006  & $-$85 &   0.994 &    0.25  &    0.02  &    0.27&   2.15  \cite{Smith:1911} \\\hline
 FePt       & 293& 52566& 35.90  & -12.40   &   828& 2.007  &   132 &   0.957 &    4.61  &    1.27  &    5.88&   5.60  \cite{Mizuguchi:2012}\\\hline\hline
Fe   &\multicolumn{12}{l}{$4\pi M_{z}=22100$ G\cite{Butler:1940}} \\
Co   &\multicolumn{12}{l}{$4\pi M_{z}=17857$ G\cite{Stifler:1911}} \\
FePt &\multicolumn{12}{l}{$4\pi M_{z}=12566$ G\cite{Mizuguchi:2012}} \\
\end{tabular*}
\end{table}

\clearpage

\begin{table}
\caption{The same as in \tab{tab:Exp} for Ni calculated within the GGA+$U$ approach. 
The values of the parameters $U$ and $J$ are given in eV.}
\label{tab:Exp2}
\begin{tabular*}{\textwidth}{@{\extracolsep{\fill}}lrrdd|rrrd>{\cG\bgroup}r<{\egroup}>{\cB\bgroup}r<{\egroup}>{\cR\bgroup}r<{\egroup}|d}
\hline\hline
  & \multicolumn{1}{c}{$T$}
  & \multicolumn{1}{c}{$B$}
  & \multicolumn{1}{c}{$\rho$} 
  & \multicolumn{1}{c|}{$S$} 
  & \multicolumn{1}{c}{$\sigmaic$} 
  & \multicolumn{1}{c}{$\alphaic/T$} 
  & \multicolumn{1}{c}{$\sigmasj$} 
  & \multicolumn{1}{c}{$\alphasj/T$} 
  & \multicolumn{1}{c}{\cG{$Q_{1}^{\q{ic}}$}}
  & \multicolumn{1}{c}{\cB{$Q_{1}^{\q{sj}}$}}
  & \multicolumn{1}{c|}{\cR{$Q_{1}^{\q{ic+sj}}$}}    
  & \multicolumn{1}{c}{$Q_{1}$ Expt.}    
   \\ \hline
 Ni$^{U=0.0}_{J=0.0}$ 
            & 291&  6290& 7.03^*&-18.75^* &$-$2150 &$-$9.057 & 350 & -4.169 & $-$8.27 &$-$0.69 & $-$8.96& 3.55  \cite{Zahn:1904}  \\    
            & 291& 10620& 7.03^*&-18.75^* &$-$2150 &$-$9.057 & 350 & -4.169 & $-$8.27 &$-$0.69 & $-$8.96& 1.30  \cite{Zahn:1904}  \\    
            & 300&      & 7.42^*&-19.19^* &$-$2140 &$-$8.482 & 349 & -4.034 & $-$8.71 &$-$0.71 & $-$9.42& 5.25  \cite{Smith:1911} \\    
            & 311&  9000& 7.90^*&-19.73^* &$-$2128 &$-$7.814 & 347 & -3.873 & $-$9.23 &$-$0.72 & $-$9.96& 2.59  \cite{Hall:1925}  \\    
            & 313&  6670& 12.15 &-19.83^* &$-$2126 &$-$7.697 & 347 & -3.844 &$-$14.19 &$-$1.10 &$-$15.30& 5.93  \cite{Butler:1940}\\
            & 323&  6670& 12.63 &-20.32^* &$-$2115 &$-$7.129 & 345 & -3.702 &$-$14.70 &$-$1.10 &$-$15.80& 6.62  \cite{Butler:1940}\\
            & 330&  9000& 8.72^*&-20.66^* &$-$2107 &$-$6.750 & 344 & -3.604 &$-$10.12 &$-$0.74 &$-$10.86& 3.04  \cite{Hall:1925}  \\
            & 333&  6670& 13.04 &-20.81^* &$-$2104 &$-$6.591 & 343 & -3.563 &$-$15.12 &$-$1.09 &$-$16.20& 7.31  \cite{Butler:1940}\\
            & 343&  6670& 13.46 &-21.30^* &$-$2092 &$-$6.081 & 342 & -3.428 &$-$15.53 &$-$1.06 &$-$16.59& 8.00  \cite{Butler:1940}\\
            & 350&  9000& 9.59^*&-21.64^* &$-$2085 &$-$5.740 & 340 & -3.336 &$-$11.03 &$-$0.73 &$-$11.76& 3.66  \cite{Hall:1925}  \\
            & 350&      & 9.59^*&-21.64^* &$-$2085 &$-$5.740 & 340 & -3.336 &$-$11.03 &$-$0.73 &$-$11.76& 7.25  \cite{Smith:1911} \\\hline 
 Ni$^{U=1.9}_{J=1.1}$ 
            & 291&  6290& 7.03^*&-18.75^* &$-$1112 &   0.999 & 409 & -0.111 & $-$2.23 &  0.91  & $-$1.31& 3.55  \cite{Zahn:1904}  \\       
            & 291& 10620& 7.03^*&-18.75^* &$-$1112 &   0.999 & 409 & -0.111 & $-$2.23 &  0.91  & $-$1.31& 1.30  \cite{Zahn:1904}  \\       
            & 300&      & 7.42^*&-19.19^* &$-$1112 &   1.047 & 409 & -0.078 & $-$2.38 &  1.00  & $-$1.39& 5.25  \cite{Smith:1911} \\       
            & 311&  9000& 7.90^*&-19.73^* &$-$1113 &   1.105 & 408 & -0.039 & $-$2.58 &  1.10  & $-$1.48& 2.59  \cite{Hall:1925}  \\       
            & 313&  6670& 12.15 &-19.83^* &$-$1113 &   1.115 & 408 & -0.032 & $-$3.98 &  1.71  & $-$2.27& 5.93  \cite{Butler:1940}\\       
            & 323&  6670& 12.63 &-20.32^* &$-$1114 &   1.166 & 408 &  0.003 & $-$4.20 &  1.85  & $-$2.35& 6.62  \cite{Butler:1940}\\       
            & 330&  9000& 8.72^*&-20.66^* &$-$1114 &   1.200 & 408 &  0.027 & $-$2.93 &  1.31  & $-$1.62& 3.04  \cite{Hall:1925}  \\       
            & 333&  6670& 13.04 &-20.81^* &$-$1114 &   1.215 & 408 &  0.037 & $-$4.40 &  1.98  & $-$2.42& 7.31  \cite{Butler:1940}\\       
            & 343&  6670& 13.46 &-21.30^* &$-$1115 &   1.262 & 408 &  0.070 & $-$4.61 &  2.12  & $-$2.49& 8.00  \cite{Butler:1940}\\       
            & 350&  9000& 9.59^*&-21.64^* &$-$1115 &   1.294 & 408 &  0.092 & $-$3.32 &  1.55  & $-$1.77& 3.66  \cite{Hall:1925}  \\       
            & 350&      & 9.59^*&-21.64^* &$-$1115 &   1.294 & 408 &  0.092 & $-$3.32 &  1.55  & $-$1.77& 7.25  \cite{Smith:1911} \\\hline 
 Ni$^{U=3.9}_{J=1.1}$ 
            & 291&  6290& 7.03^*&-18.75^* &$-$801  &   4.627 & 430 & -0.904 & $-$0.19 &  0.67  &    0.48& 3.55  \cite{Zahn:1904}  \\      
            & 291& 10620& 7.03^*&-18.75^* &$-$801  &   4.627 & 430 & -0.904 & $-$0.19 &  0.67  &    0.48& 1.30  \cite{Zahn:1904}  \\      
            & 300&  --- & 7.42^*&-19.19^* &$-$802  &   4.637 & 429 & -0.863 & $-$0.19 &  0.74  &    0.55& 5.25  \cite{Smith:1911} \\      
            & 311&  9000& 7.90^*&-19.73^* &$-$802  &   4.647 & 428 & -0.815 & $-$0.19 &  0.82  &    0.63& 2.59  \cite{Hall:1925}  \\      
            & 313&  6670& 12.15 &-19.83^* &$-$802  &   4.648 & 427 & -0.806 & $-$0.29 &  1.27  &    0.98& 5.93  \cite{Butler:1940}\\      
            & 323&  6670& 12.63 &-20.32^* &$-$803  &   4.653 & 426 & -0.765 & $-$0.29 &  1.38  &    1.09& 6.62  \cite{Butler:1940}\\      
            & 330&  9000& 8.72^*&-20.66^* &$-$803  &   4.656 & 426 & -0.737 & $-$0.19 &  0.98  &    0.79& 3.04  \cite{Hall:1925}  \\      
            & 333&  6670& 13.04 &-20.81^* &$-$803  &   4.656 & 425 & -0.725 & $-$0.28 &  1.48  &    1.20& 7.31  \cite{Butler:1940}\\      
            & 343&  6670& 13.46 &-21.30^* &$-$804  &   4.656 & 424 & -0.687 & $-$0.27 &  1.58  &    1.31& 8.00  \cite{Butler:1940}\\      
            & 350&  9000& 9.59^*&-21.64^* &$-$804  &   4.655 & 424 & -0.661 & $-$0.19 &  1.16  &    0.97& 3.66  \cite{Hall:1925}  \\      
            & 350&  --- & 9.59^*&-21.64^* &$-$804  &   4.655 & 424 & -0.661 & $-$0.19 &  1.16  &    0.97& 7.25  \cite{Smith:1911} \\\hline
 Ni$^{U=3.9}_{J=2.6}$ 
            & 291&  6290& 7.03^*& -18.75^* &$-$842& 6.347 & 444 &  -0.379 &   0.33 &  0.90 &  1.23& 3.55  \cite{Zahn:1904}  \\      
            & 291& 10620& 7.03^*& -18.75^* &$-$842& 6.347 & 444 &  -0.379 &   0.33 &  0.90 &  1.23& 1.30  \cite{Zahn:1904}  \\      
            & 300&  --- & 7.42^*& -19.19^* &$-$843& 6.330 & 443 &  -0.355 &   0.37 &  0.97 &  1.34& 5.25  \cite{Smith:1911} \\      
            & 311&  9000& 7.90^*& -19.73^* &$-$844& 6.307 & 443 &  -0.327 &   0.41 &  1.08 &  1.49& 2.59  \cite{Hall:1925}  \\      
            & 313&  6670& 12.15 & -19.83^* &$-$844& 6.302 & 442 &  -0.322 &   0.64 &  1.66 &  2.30& 5.93  \cite{Butler:1940}\\      
            & 323&  6670& 12.63 & -20.32^* &$-$845& 6.276 & 442 &  -0.298 &   0.69 &  1.79 &  2.48& 6.62  \cite{Butler:1940}\\      
            & 330&  9000& 8.72^*& -20.66^* &$-$846& 6.257 & 441 &  -0.282 &   0.49 &  1.26 &  1.75& 3.04  \cite{Hall:1925}  \\      
            & 333&  6670& 13.04 & -20.81^* &$-$846& 6.248 & 441 &  -0.275 &   0.74 &  1.90 &  2.64& 7.31  \cite{Butler:1940}\\      
            & 343&  6670& 13.46 & -21.30^* &$-$847& 6.218 & 440 &  -0.253 &   0.78 &  2.02 &  2.80& 8.00  \cite{Butler:1940}\\      
            & 350&  9000& 9.59^*& -21.64^* &$-$847& 6.195 & 439 &  -0.237 &   0.57 &  1.47 &  2.03& 3.66  \cite{Hall:1925}  \\      
            & 350&  --- & 9.59^*& -21.64^* &$-$847& 6.195 & 439 &  -0.237 &   0.57 &  1.47 &  2.03& 7.25  \cite{Smith:1911} \\\hline\hline
    \multicolumn{13}{l}{$4\pi M_{z}=\phantom{0}5670$ G\cite{Butler:1940}} \\
\end{tabular*}
\end{table}

\clearpage

%